\documentclass[twocol]{ametsocV5}
\usepackage{amsmath,amsfonts,amssymb,bm}
\usepackage{mathptmx}
\usepackage{newtxmath}

\usepackage{graphicx}
\usepackage{dcolumn}
\pdfoutput=1
\usepackage{multicol, blindtext}
\newcommand\nc\newcommand
\def\fou#1{\widehat{#1}} 
\nc\vect[1]{\boldsymbol{#1}}
\nc\be{\begin{equation}}
\nc\ee{\end{equation}}
\def\abs#1{\left\vert #1 \right\vert} 
\def\norme#1{\left\Vert #1 \right\Vert} 
\newcommand\Real{\mbox{Re}} 
\newcommand\Imag{\mbox{Im}} 
\def\cali#1{\cal#1\mit}

 \authors{Charles-Antoine Gu\'erin\correspondingauthor{Charles-Antoine Gu\'erin, guerin@univ-tln.fr},
   Dylan Dumas, Anne Molcard, \\C\'eline Quentin, Bruno Zakardjian, Anthony Gramoull\'e}
\affiliation{ MIO, Universit\'e de Toulon, Aix-Marseille Univ, CNRS, IRD, Toulon, France}
\extraauthor{ Maristella Berta}
\extraaffil{ISMAR, CNR, La Spezia, Italy}

\title{High-Frequency radar measurements with CODAR in the region of Nice: improved calibration and performance}

\abstract{ We report on the installation and first results of one compact oceanographic radar in the region of Nice for a long-term observation of the coastal surface currents in the North-West Mediterranean Sea. We describe the specific processing and calibration techniques that were developed at the laboratory to produce high-quality radial surface current maps. In particular, we propose an original self-calibration technique of the antenna patterns, which is based on the sole analysis of the database and does not require any shipborne transponder or other external transmitters. The relevance of the self-calibration technique and the accuracy of inverted surface currents have been assessed with the launch of 40 drifters that remained under the radar coverage for about 10 days.}
\begin{document}

\maketitle

\section{Introduction}
In recent years there has been a growing European effort for the integrated observation of the Mediterranean Sea (e.g. \cite{bellomo_toward_2015}), one of the current challenges being the monitoring of the oceanic circulation in the region. In that respect, land-based High-Frequency Radars (HFR) have proven for a few decades to be very efficient instruments to measure ocean surface currents over the clock over broad coastal areas (e.g. \cite{paduan_AnnRev13}). This has motivated the ongoing construction of a spatially and temporally dense HFR network for a continuous coverage of the Mediterranean coasts \cite{quentin_hf_2013}. In the framework of the long-term French observation program ``Mediterranean Ocean Observing System for the Environment'' (\cite{MOOSE,coppola_EOS19}) and more recently the EU Interreg Marittimo  ``Sicomar-Plus'' project (\cite{sicomar_sistema_2018}), our laboratory has installed a pair of HFR of SeaSonde type (CODAR Ocean Sensors, Ltd.) in the region of Nice. One of them has been operational since 2013 in Saint-Jean Cap Ferrat,  2 km East of Nice. The other has been recently installed at the French-Italian border in the Port of Menton and is currently being evaluated.

CODARs are commonly used instruments in the oceanographic community (e.g. \cite{lipa_JOE06}); they have a dedicated software that in principle does not require any specific skills in the field of radar instrumentation and signal processing. However,  we found that the standard performances proposed by the commercial toolbox suite can be greatly improved by using a specific home-made processing, which can compensate for some defects of the installation.  {\color{black} The main limitations of such radars are related to their weak azimuthal resolution comparing to other radars (e.g. \cite{Liu_JAOT14}) and the touchy calibration of their complex antenna gains; these shortcomings are the counterpart of the compactness of the antenna system.} As it is well known, the accurate characterization of the antenna patterns is a key point in the Direction Finding (DF) technique that is employed to obtain the azimuthal discrimination of radar echoes with compact HFR systems. The antenna system in real conditions undergoes in general some distortion with respect to the factory settings and a precise on-site calibration of the individual antenna responses is required before proceeding to surface current mapping (\cite{paduan_JOE06}).

In this report we describe the first operational HFR site of  Saint-Jean Cap Ferrat and its specificity (Section 2) and we recall the principle of surface current mapping in this context. The key point is the appropriate utilization of DF algorithm to resolve the surface current in azimuth, a process that goes through the application of the MUSIC algorithm (Section 3). We present some techniques from the user point of view that have an important impact on the quality of the result. We describe a recent ship experiment to measure the real antenna pattern following the standard procedure and we give a detailed review of the involved radar signal processing steps (Section 4). In Section 5, we propose a novel self-calibration technique based on the mere analysis of the recorded Doppler spectra. This method does not require any specific calibration campaign and allows an easy update of the calibration as time goes by. We compare the radial surface current maps obtained with 3 types of antenna pattern (ideal, measured, self-calibrated) and assess them in the light of a recent oceanographic campaign providing the {\it in situ} measurements of surface current by means of a set of 40 drifters (Section 6). We show that the best agreement is obtained with the self-calibrated antenna patterns while the improvement brought by the measured pattern over the ideal is marginal in that case.

\section{Radar sites history and description}

The installation of one CODAR SeaSonde sensor in the region of Nice was initially driven by the need to monitor the local oceanic circulation in the framework of the MOOSE network \cite{quentin_moose14}. The main objective was to properly track the Northern Current (NC), which is the dominant stream flowing along the continental slope. It transports warm and salty Levantine Intermediate Water at intermediate depths and thereby affects the properties of the Western Mediterranean Deep Water formed downstream off the Gulf of Lion. These radar sites are complementary to another site located in the region of Toulon downstream the NC (\cite{dumas_multistatic_2020}). The choice of this instrument was mainly motivated by its compactness that is an absolute requirement in the densely urbanized coastal zone of the French riviera. The HFR station was initially meant to cover the geographical area surrounding a mooring site referred to as ``DYFAMED'' for the long-term observation of the core parameters of the water column (\cite{coppola_PO18,coppola_EOS19}). The combination of the time series describing the vertical structure and the HFR currents is essential for the characterization of the full 3D structures of circulation observed in the area (\cite{berta_OS18,coppola_PO18,manso_OS20}). This mooring site is located 50 km offshore and a medium range HFR (9-20 MHz operating frequency) is well adapted to cover the surface current measurements from the coast to 70-80 km away.

\begin{figure}[htbp]\centering
  \includegraphics[scale=0.55]{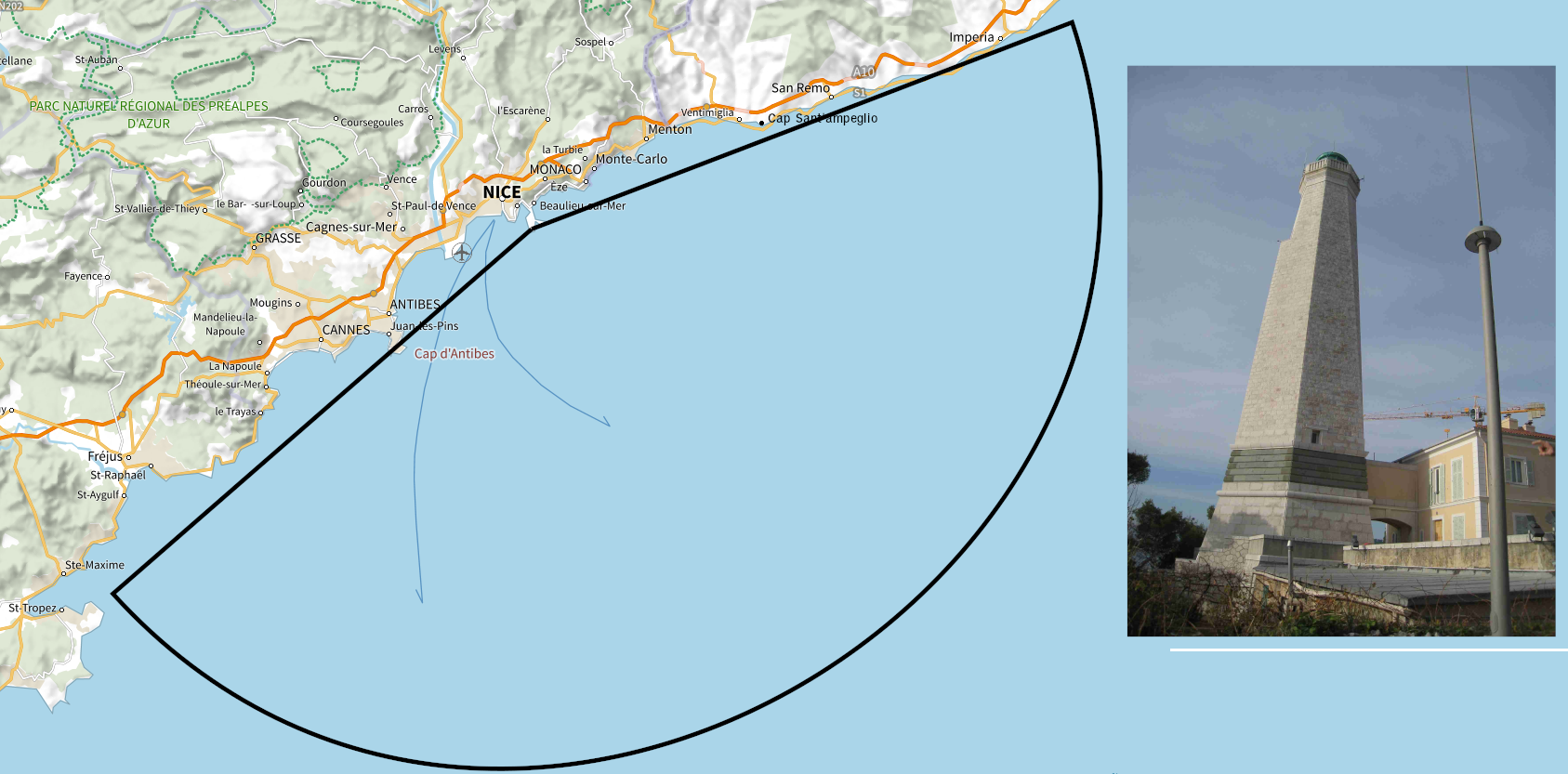}
  
        \caption{Location and theoretical coverage of the SeaSonde CODAR installed in Saint-Jean Cap Ferrat. The field of view is limited to the West by the Cape of Antibes and to the East by Cape Saint Ampelio. The arc of circle marks a typical 65 km coverage. The SeaSonde mast at the lighthouse of Saint-Jean Cap Ferrat (upper right).\label{figCODAR}}
\end{figure}

The SeaSonde HFR operates at a 13.5 MHz central frequency within a full bandwidth of 100 kHz allowing for 1.5 km range resolution. It is sited at the lighthouse of Saint-Jean Cap Ferrat where it is installed on a flat terrace less than 90 m away from the sea at an altitude of about 30 meter (Figure \ref{figCODAR}). It is equipped with co-located receive and transmit antennas mounted on a 7 meter mast. It is separated by only 28 m from the 34 m high light house tower that shields a small angular sector to the West. Electronic devices including transmitter and receiver as well as the computer device (under MacOS system) controlling the equipment are hosted in an air-conditioned enclosure as recommended (\cite{mantovani2020best}).

Albeit far from ideal, this configuration was found the best compromise given the field constraints. The magnetic compass direction of the receive antenna sighting arrow is about 160 degrees corresponding to a theoretical bearing of 207 degree true North Clock Wise (NCW) after correction of the magnetic declination. The geometry of the coast between the Cape of Antibes on the Western side and Cape Martin on the Eastern side allows viewing an angular sector from $60^\circ$ to $225^\circ$ NCW but the distance between the SeaSonde and the sea remains constant (at about 50 m) only from $135^\circ$ to $315^\circ$ NCW. The vicinity of the lighthouse tower and the increased ground distance are expected to limit the performances of the radar in the other azimuthal directions.

\section{Radial current extraction with CODARs\label{sec:DoA}}

As it relies on a compact antenna array, the CODAR uses DF algorithms to resolve the surface current in azimuth, which is the most critical step in the processing chain of the radar data and requires a reliable knowledge of the complex antenna gains. We recall that the antenna system is composed of two cross-loop antennas and one monopole, with ideal patterns:
\be
\begin{split}
  g_1(\theta)&=\cos(\theta-\theta_b),\\
  g_2(\theta)&=\sin(\theta-\theta_b),\\
  g_3(\theta)&=1
  \end{split}
\ee
where $\theta_b$ is the main bearing. In its standard mode the CODAR SeaSonde uses Interrupted Frequency Modulated Continuous Wave with chirps of duration $0.25$ sec {\color{black}followed by a blank of same duration $0.25$ sec} leading to a 2 Hz sampling rate for the range-resolved backscattered temporal signal $X_j$ on each antenna. A Fast Fourier Transform is applied to these time series to form the co- and cross-Doppler spectra $D_{ij}=\fou X_i\fou X_j^\ast$ (9 elements), where $\fou X_j$ denotes the Discrete Fourier Transform and $\ast$ the complex conjugate. A typical integration time of 256 sec (512 chirps) is employed to obtain the individual complex Doppler spectra $D_{ij}$ (512 Doppler bins, ``CSQ'' files), which are averaged over 15 min (4 CSQ) to produce mean Doppler spectra $\overline D_{ij}$ at an update rate of 10 min (``CSS'' files). A series of 7 such spectra are further combined to produce an ensemble average over 75 min (with update rate of 60 min) and estimate a covariance matrix $\Sigma_{ij}=\langle\overline D_{ij}\rangle$ for each Doppler bin and each range bin. The Direction of Arrival (DoA) corresponding to a given frequency shift (hence to a given radial surface current) is determined from the MUSIC algorithm (\cite{schmidtAP86}). A Singular Value Decomposition (SVD) of each covariance matrix is performed, with eigenvalues $\lambda_1\geq\lambda_2\geq\lambda_3\geq 0$ and normalized eigenvectors $\vect V_1,\vect V_2,\vect V_3$. In the MUSIC theory, the number of nonzero eigenvalues corresponds to the number of sources (i.e, the number of occurrences of a given radial current when sweeping over azimuth) and the associated eigenvectors span the signal subspace; the remaining null eigenvalues span the noise space. {\color{black} In practice, the noise eigenvalues are not truly null as in an ideal system and a threshold must be selected to distinguish the signal and noise eigenvalues.} As the antenna array of the CODAR SeaSonde is limited to 3 elements, the MUSIC algorithm can be applied with at most two sources to preserve at least one dimension for the noise subspace. As the electromagnetic field incoming from any far source is multiplied by the complex antenna gain, the so-called steering vector $\vect g(\theta)= (g_1(\theta), g_2(\theta),g_3(\theta))$ in the corresponding direction turns out to be an eigenvector of the covariance matrix and spans its signal subspace. Hence, the DoA attached to any source can be obtained by maximizing the projection of the steering vector on the signal subspace or equivalently by minimizing its projection onto the noise subspace. High-resolution estimate of the DoA are rather obtained by forming the ``MUSIC Factor'', which maximizes the inverse projection of the steering vector on the noise space. In Mathematical terms, the detected bearing $\theta_{DoA}$ is given by $\theta_{DoA}=\mathrm{Argmax}\ Q(\theta)$, with
\be
  Q(\theta)=\left( \frac{\norme{\vect g(\theta)}^2}{\norme{\sum_{n=3-N_s+1}^3 (\vect V_n\cdot \vect g(\theta)) \vect V_n}^2}\right)
\ee
where $N_s=1$ or $2$ is the number of sources. In practice, the quality of the MUSIC estimation relies on the appropriate setting of a certain number of tuning parameters such as the number of sources, the angular bin, the different thresholds used for the selection of Bragg frequency bin and for the MUSIC factors etc (see e.g. \cite{kirincich_JAOT17,kirincich_JAOT19,emery_JAOT19}. In addition, we found some extra numerical recipes that we detail here.

\begin{figure}[htbp]
\begin{minipage}[c]{0.95\textwidth}
\hspace{-1.25 cm} \includegraphics[scale=0.27]{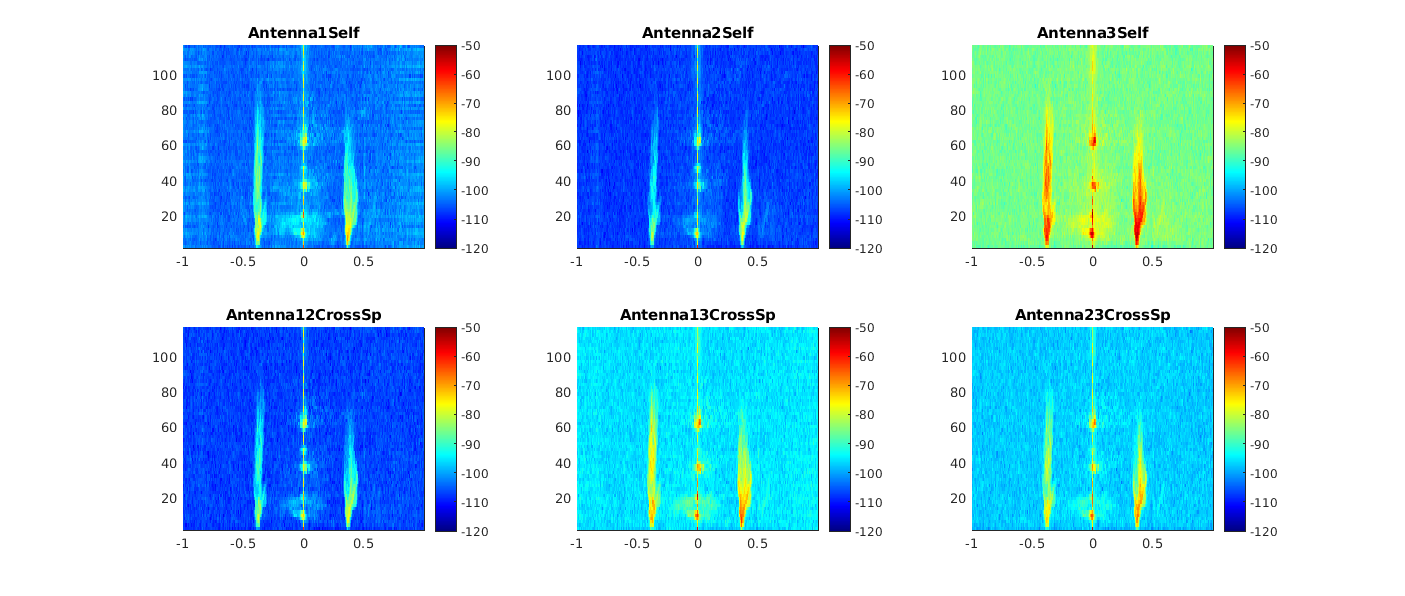} \\a)
\end{minipage} \begin{minipage}[c]{0.95\textwidth}
\hspace{-1.25 cm}  \includegraphics[scale=0.27]{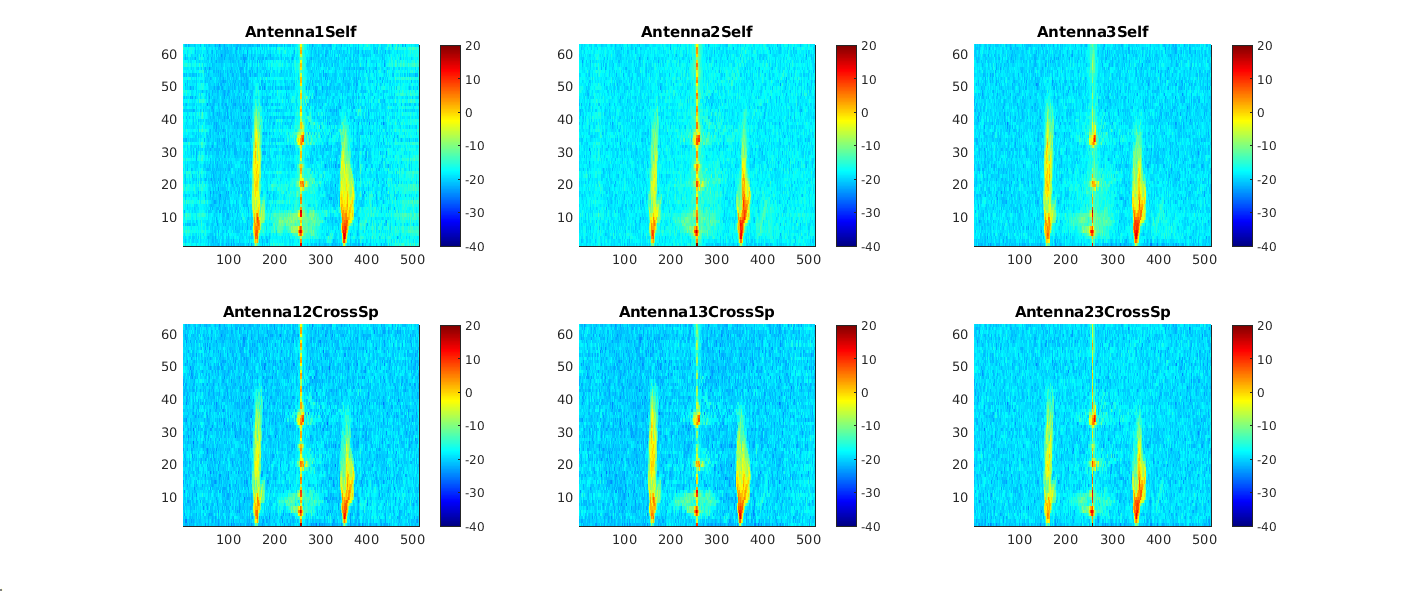}\\ b) \end{minipage} 
 \caption{Examples of hourly Doppler co- and cross-spectra before (a) and after (b) normalization.\label{FIG0AB}}
\end{figure}

\begin{figure}[htbp]
\hspace{-1cm}  \includegraphics[scale=0.4]{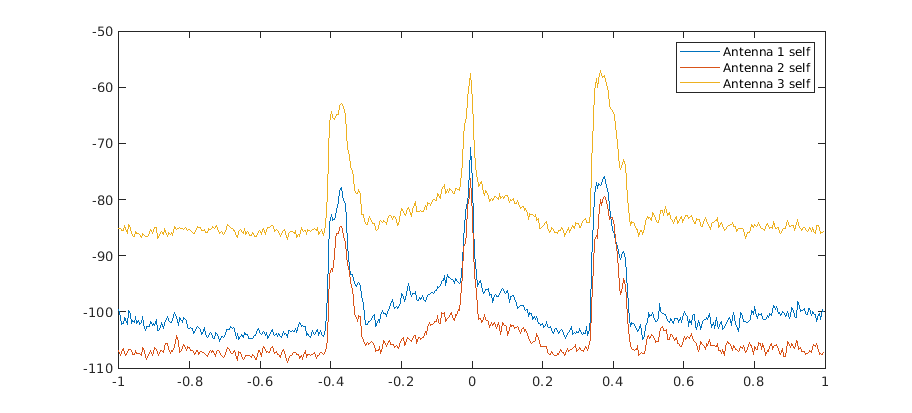}
 \caption{A ``slice'' of the above Doppler co-spectra averaged over ranges from 5 to 10 km. As seen, the 3 antennas do not have the same absolute gain.\label{FIG0C}}
\end{figure}

\paragraph{Normalization of the covariance matrix}
The covariance matrix, because it is obtained from a nonlinear combination of the antenna channels, is strongly dependent on the individual calibration of the signals recorded on each antenna. {\color{black} This is illustrated by the Doppler co-spectra shown in Figure \ref{FIG0C} for each antenna, where an important shift of the level of noise floor and Bragg peaks is observed.} Now, rescaling one channel in amplitude will strongly affects the eigenvalues and eigenvectors of the covariance matrix, hence the whole DF process. This cannot be compensated by a simple rescaling of the steering vector. It is therefore important that the diagonal coefficients of the covariance matrix have the same order of magnitude. This implies that the Doppler co-spectra should also have the same order of magnitude from the beginning, aside from angular variations due to the antenna pattern. For this it is not enough to form the ``loop amplitude ratios'', which are the signals received on antenna loop 1 and 2 normalized by the reference signal received on the monopole antenna 3, because the gains of antenna 1 and 2 might have very different amplitudes. Instead, we consider the complex Signal-to-Noise Ratio (SNR), which can be obtained by normalizing the complex Bragg spectra lines by the background noise level (defined as the average Doppler spectrum amplitude in the far range away from the Bragg lines). This amounts to renormalize the covariance matrix by the noise levels:
  \be
\Sigma_{ij}\mapsto \Sigma_{ij}/(\sqrt{N_iN_j}),
  \ee
  where $N_i$ is the noise level attached to the $i$th antenna {\color{black} and $\Sigma_{ij}$ are the elements of the covariance matrix. This operation has been found to improve greatly the consistency between the matrix eigenvectors and the ideal steering vectors. The 6 different elements of the covariance matrix (3 co- and 3 cross-spectra) are shown in Figure \ref{FIG0AB} before and after renormalization with this technique. This operation allows to restore a same order of magnitude for the 6 elements.}

\paragraph{ {\color{black}Consistency of negative and positive Bragg lines}}

{\color{black} A customary quality check when estimating the position of the Bragg lines is to verify that they undergo the same shift
  due to the translational effect of surface current. This implies that the difference of the positive and negative shifted Bragg frequency should be equal
  to twice the Bragg frequency, $f_B^+-f_B^-=2 f_B$. However, this criterion can only be used as such when using beam-forming techniques where the Doppler spectrum is
  azimuthally resolved. Nevertheless, a consistency criterion can still be devised with DF techniques by comparing the estimations of the surface currents $U_r$ obtained for each range and each angular bin
  with either the positive ($U^+_r$) or the negative Doppler frequencies ($U_r^-$), providing that the corresponding Bragg lines have sufficient SNR. This has been used as an extra quality control to validate the estimated values of the surface current. In the presented numerical calculations we required that $\abs{U^+-U^-}<0.15$ m/s.}

\paragraph{Temporal stacking}

A known shortcoming of the MUSIC method for the azimuthal discrimination is the lacunarity of the resulting radial current maps. This default can be mitigated by the method of antenna grouping when a sufficient number of antennas is available (\cite{dumas_self-calibration_2020}). This technique cannot be applied to the CODAR that has only 3 antennas. However, the idea of applying the DoA algorithm on different groups to increase the coverage has been retained and adapted to the SeaSonde case. The idea is to distinguish groups of antenna in time instead of space, at the cost of an increased observation time. For this, we consider all different groups of sample spectra (CSS) that can be used to estimate the covariance matrix. Precisely, rather than applying the MUSIC algorithm to a single group of 7 CSS within one hour record, we also apply it to other available incomplete groups of samples. There are 2 possible groups of 6 consecutive CSS, 3 groups of 5 CSS, etc. This amounts to apply the DoA algorithm as many times as there are admissible groups, thereby obtaining multiple estimates of the spatial distribution of surface current. In the end, a radial current map with a significantly increased coverage can be obtained by averaging the individual maps derived from every single group. Starting with a minimal size of 3 CSS, there are 15 admissible groups of consecutive CSS. Note that smaller groups provide coarser estimates of the covariance matrix hence the radial current direction. However, they help filling the map while higher-order groups ensure more accurate estimates. Different weights can be attributed to the latter to favor large groups and increase the accuracy. We refer to this technique as ``temporal stacking'', as 'stacking' is a well-known technique in {\color{black}astrophotography to reduce the noise of short exposure photography by accumulating redundant frames (e.g. \cite{schroder2009astrophotography}). It is difficult to evaluate the accuracy of temporal stacking {\it per se}. In principle, resorting to smaller groups should deteriorate the global accuracy of the DoA. However, increasing the number of available current values falling within the same directional bin allows an easy test to eliminate untrustworthy estimates, by rejecting those directions that show inconsistent distribution of values (for example, having an unacceptably large standard deviation). By using such control criterion (here with a maximum standard deviation of 20 cm/s), we could improve the accuracy of the stacking method while augmenting the radar coverage (see Figure \ref{cartes3methodes}d). This has been confirmed by our comparisons of HFR radial currents with drifter measurements (section \ref{sec:drifter}) which show that  using temporal stacking does cause any visible change in the statistical performances (see Table \ref{tablestacking}) while increasing the number of sample points.}

\section{Ship calibration of the antenna patterns}

\subsection{Standard calibration procedure}
As already mentioned the antenna calibration is a crucial step for the quality of HFR surface current measurement (e.g. \cite{paduan_JOE06}). The CODAR SeaSonde manufacturer provides a specific transponder and a well-established procedure to perform this calibration with a shipborne transmission. The transponder interacts with the radar receiver to characterize the antenna pattern in the practical configuration of the field deployment. The boat describes a circle around the antennas while the HFR system acquires the transponder signal. The exact ship position is tracked with a GPS device and the track must be traveled clockwise and counterclockwise to eliminate some possible bias in time and average the measurements.

\begin{figure}[htbp]\centering
  \includegraphics[scale=1]{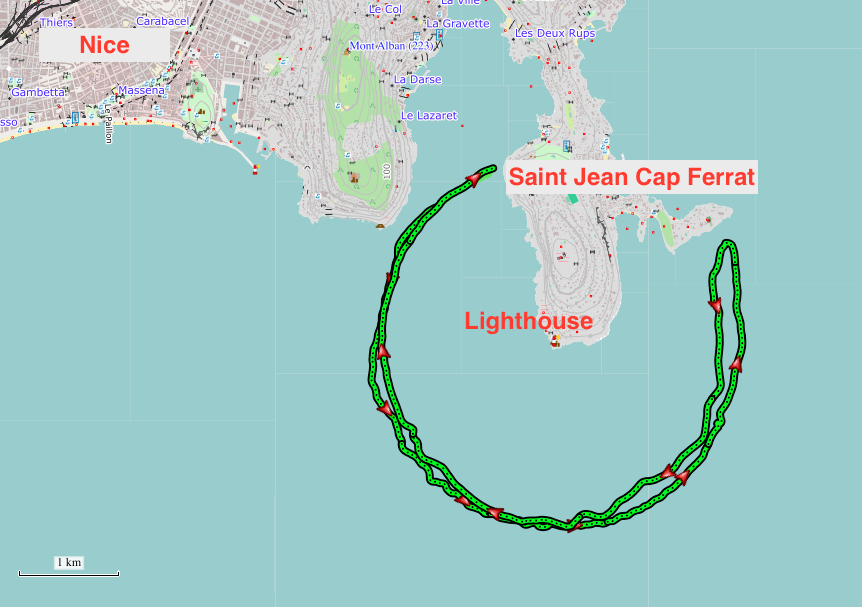}
  
  \caption{Ship trajectory for the antenna pattern campaign\label{shiptrajectory}}
\end{figure}

The field campaign for the antenna calibration was carried over on March, 11, 2020 with a standard SeaSonde transponder mounted on the Pelagia boat (Institut de la Mer de Villefranche). The distance was maintained constant at one nautical mile from the SeaSonde antenna, a short range that is imposed by the weak power of the transponder. At this distance the geographic situation (peninsula of Cap Ferrat) allows to cover a circular arc from $60^\circ$ to $330^\circ$ NCW. The ship trajectory is shown on Figure \ref{shiptrajectory}. The boat was driven at a speed ranging from 10 km/h to 15 km/h that is equivalent to an angular velocity between 4 to 6 degree of arc per minute, well below the 8.5 knots (15.7 km/h) velocity threshold recommended by experienced users (\cite{APMguide,Washburn_JAOT17}) for such measurements.

\begin{figure}[htbp]\centering
  \includegraphics[scale=1.4]{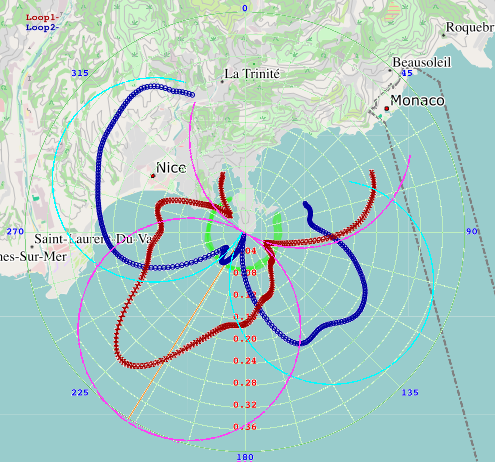}
  
        \caption{Experimental antenna patterns derived from the SeaSonde software. The red (blue) thick solid lines show the smoothed measured angular pattern. The magenta (cyan) thin solid lines indicate the ideal pattern at the main bearing extrapolated from the minimum of loop 2 (orange solid line).\label{seasondepattern}}
\end{figure}

The recorded time series were processed by the SeaSonde software to extract the antenna pattern and were also processed from scratch in the laboratory. We recall hereafter the main steps of the calibration.\\

\noindent 1. Record the raw I/Q time series (Llv files) on the 3 receiver channels when the transmitter is a shipborne transponder.\\
2. Process the received signal in range (fast time) and Doppler (slow time) to obtain Range-Doppler maps. Thanks to the GPS track of the ship, every such map can be associated to an azimuth $\theta$.\\
3. Identify the bright spot corresponding to the ship echo in the Range-Doppler domain and the corresponding Range-Doppler cell. The transponder and the receiver parameters have been set in order to locate a signal peak in range cells from 10 to 13 and Doppler cells from 35 to 55. \\
4. For each antenna, calculate the complex gain $S_1(\theta),S_2(\theta),S_3(\theta)$ of each antenna in the direction of the ship by evaluating the complex Doppler spectrum in the related cell. Ideally, the ship response involves a single cell in the Range-Doppler map. In practice, the ship illuminates a few neighbor cells. The antenna gain can be chosen as the mean complex Doppler spectrum over these neighbor cells or the complex value where the amplitude is maximum. The so-called complex loop amplitude ratios of antennas 1 and 2 are obtained by normalizing the ship response by antenna 3 (monopole), that is $g_1(\theta)=S_1/S_3$ and $g_2(\theta)=S_2/S_3$.\\
5. Perform a sufficient angular smoothing of the complex loop amplitude ratios to remove the fast oscillations.\\

Figure \ref{seasondepattern} shows the antenna power patterns (that is, $\abs{g_1(\theta)}^2$ and $\abs{g_2(\theta)}^2$) obtained with the SeaSonde software processing of the field measurement after an angular smoothing over a 10 degrees window and interpolated with 1 degree steps. A significant deviation from the ideal pattern can be observed both in shape and orientation. The main bearing cannot be determined without ambiguity as the maximum of loop 1 does not coincide with the minimum of loop 2. The position of the latter is, however, difficult to estimate as it is hidden by a secondary lobe. Based on the extrapolation of the edges of the main 2 lobes of loop 2, the main bearing is found at 212 degree NCW (versus 207 degree when measured with the magnetic compass and the mast marker). This bearing value will be used in the following to characterize the ideal pattern of the Ferrat station.

\subsection{Laboratory calibration}

\begin{figure}[htbp]\centering
\hspace{-0.75cm} \includegraphics[scale=0.5]{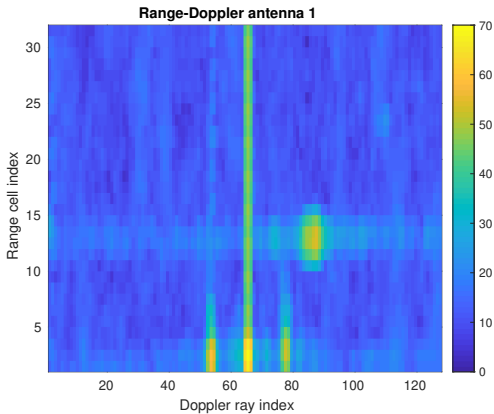}
  
          	\caption{Example of Range-Doppler spectra on the three antennas (in dB scale). The bright spot around coordinates 90-12
                is due to the ship motion. The zero Doppler bin (central vertical line) as well as the Bragg lines are clearly visible.\label{RD}}
\end{figure}

The quality of the overall pattern processing is strongly dependent of the windowing process and the integration times that are chosen to perform the range resolution and the Doppler analysis. Each Llv file consists of 512 sweeps corresponding to total duration of 256 second. Within each sweep (duration = 0.5 sec), the received I/Q signal is sampled at a fast time with 4096 points. Range processing is obtained using FFT with a Blackman-Harris window. Among the 2048 available positive frequency bins, only the first 32 frequency bins are preserved, corresponding to the closest range cells from 1 to about 60 km. This provides range-resolved time series on each antenna at 2 Hz sampling frequency. Doppler spectra are obtained by applying a further FFT on these time series. A trade-off between the required integration time and the necessary update rate was found by processing 7 half-overlapping sequences of 1 minute within the 256 second duration of each Llv file, thereby providing an estimation of the complex Doppler spectrum every 30 seconds. The resulting loop amplitude ratio are further interpolated onto the GPS time scale (1 second). Figure \ref{RD} shows an example of the obtained Range-Doppler maps for the first antenna. The positive and negative Bragg lines are clearly visible together with the zero-Doppler line (fixed echoes) and a bright spot around position 90-13 corresponding to the ship signal.

The complex antenna patterns $g_1,g_2$ can be represented in amplitude and phase as:
\be
g_n(\theta)=A_n(\theta)e^{i\phi_n},\ n=1,2
\ee
In the ideal case, the amplitude $A_1,A_2$ coincide with the absolute values of the sinus and cosinus function around the main bearing and the phases $\phi_1,\phi_2$ are $0$ or $180$ degrees. The measured patterns are in the favorable case slightly distorted versions of the ideal patterns. For simplicity we assume that the distortion can be described in overall phase and amplitude but not in shape. As suggested in \cite{Laws_JAOT10} we use the following parametric form (assuming no constant bias):
\be\label{fitrhophi}
  g_1(\theta)=\rho_1\cos(\theta-\theta_{1})e^{i\phi_1},\  g_2(\theta)=\rho_2\sin(\theta-\theta_{2})e^{i\phi_2}
  \ee
for some tuning coefficients $\theta_{1},\theta_{2},\rho_1,\rho_2,\phi_1,\phi_2$. These parameters can be easily identified by fitting the real and imaginary parts of the complex loop amplitude ratios in the form:
   \be\label{fitab}
   \begin{split}
     \Real g_n(\theta)&=a_n\cos\theta+b_n\sin\theta,\ (n=1,2)\\
     \Imag g_n(\theta)&=a_n'\cos\theta+b_n'\sin\theta,\ (n=1,2)\\
     \end{split} 
   \ee
   Comparison of the expressions (\ref{fitab}) and (\ref{fitrhophi}) leads to the consistency relations:
   \be
     \begin{array}{l}
   \tan\theta_{1}=\frac{b_1}{a_1}=\frac{b_1'}{a_1'},\ \tan\theta_{2}=-\frac{a_2}{b_2}=-\frac{a_2'}{b_2'}\\
     \tan\phi_n=\frac{a_n'}{a_n}=\frac{b_n'}{b_n}, (n=1,2)\\
     \rho_n^2=a_n^2+b_n^2+(a_n')^2+(b_n')^2, (n=1,2)\
       \end{array} 
   \ee
  
\begin{table}
\center
\begin{tabular}{|c|c|c|c|c|c|c|}
  \hline
  & a & b& $\theta_0$ (deg) & $\phi$ (deg) &  $\rho$\\
  \hline
  real $g_1$& -1.6&  -2.0&  232& -34$\vert$-23 & 2.9 \\
   \hline
   imag $g_1$& 1.1 &    0.8 & 218 &  &  \\
    \hline
    real $g_2$& -0.6 &     1.4 &    203 & -50$\vert$-32 & 1.9 \\
     \hline
      imag $g_2$& 0.7 &   -0.9 &218  &  &  \\
        \hline
\end{tabular}
\caption{Estimation of the main bearing ($\theta_0$) and the amplitudes ($\rho_1,\rho_2$) and phases ($\phi_1,\phi_2$) of the complex loop amplitude ratios (\ref{fitrhophi}) from the fitting coefficients $a,b$ of equation (\ref{fitab})\label{tablefit}}
\end{table}

\begin{figure}[htbp]\centering
      \includegraphics[scale=0.3]{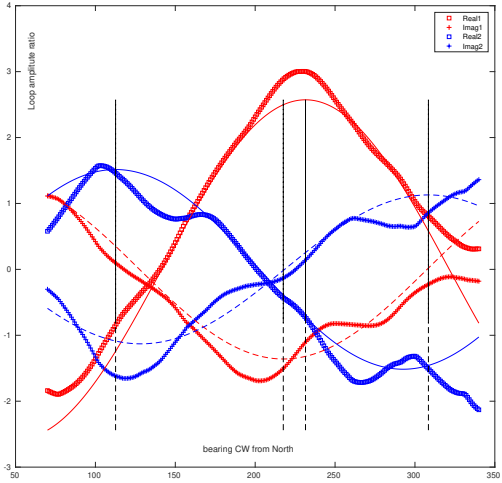}
     
        \caption{Measured (symbols) and fitted (lines) values after equation (\ref{fitab})) for the complex loop amplitude ratios (\ref{fitrhophi}). The black dashed vertical lines indicate the different values found for the estimated bearing.\label{figloopamplitude}}
\end{figure}

        The estimated values for the fitting parameters are recapped in Table \ref{tablefit}. Figure \ref{figloopamplitude} shows the real and imaginary parts of the loop amplitude ratios $g_1(\theta),g_2(\theta)$ as a function of the bearing angle together with their least-square fit (\ref{fitab}). The black dashed vertical lines indicate the different values found for the estimated bearing. A running average within a $\pm 40$ degree triangular window has been preliminary applied to smooth the measured complex patterns. By taking the average value of the parameters that have a double determination ($\theta_{1},\theta_{2},\phi_1,\phi_2$) we can estimate $\theta_{1}\simeq 225^\circ$, $\theta_{2}\simeq 210^\circ$, $\phi_1\simeq -28^\circ$ and $\phi_2\simeq -41^\circ$. As seen, the estimated angles $\theta_1$ and $\theta_2$ differ by $15$ degree, indicating that the actual patterns of loop 1 and 2 fail to satisfy the assumed quadrature relation. {\color{black} Note that the quadrature relation holds for ideal pattern but is in fact not guaranteed for distorted patterns. Hence, the difference between $\theta_1$ and $\theta_2$ is a simple indication of the level of distortion of the antennas.}

\section{Self-calibration of the antenna patterns}

\subsection{Main principle}
The experimental calibration of the antennas in real conditions is far from being satisfying both in terms of quality and tractability. The most obvious drawback is that it requires time- and money-consuming operations. Another limitation of the procedure is the strong assumption that the antenna patterns that are measured one mile away from the radar still hold at longer range, which ignores the shielding by natural remote obstacles (islands, coasts, capes,...). We have also seen that the quality of the estimation of the complex gain relies on the choice of several tuning parameters in the radar processing (length of the time series to process, tapering window, number of Doppler spectra to average, selection of the ship response in the range-Doppler maps, smoothing of the angular diagram, etc). Besides, the calibration might be subject to variations in time depending on meteorological conditions or drift of electronic devices. To avoid the heavy procedure of ship calibration, some other techniques have been proposed in the literature based on external opportunity sources (see e.g. \cite{fernandez_IGARSS03,fernandez_JOE06,flores_JAOT13,zhao_RemSen20} for antenna arrays and \cite{emery_JAOT14} for compact systems)or aerial drones \cite{Washburn_JAOT17}. We here propose a novel method that does not require any external source and is purely based on the analysis of the recorded Doppler spectra. The main idea is that a proper family of steering vector $\vect g(\theta)$ of the antenna system should cover the eigenvectors of the Doppler bins covariance matrix whenever a variety of DoA are investigated. In other words, when performing the SVD decomposition of the covariance matrix for a large number of range cells and Doppler bins and for the large variety of oceanic patterns encountered on the long-term, the obtained eigenvectors should remain within a constrained family given by the steering vector. We therefore devise the following algorithm for the estimation of the latter.

\paragraph{} We first seek a relevant representation of the antenna pattern with the help of a reduced set of parameters, $\vect X$, and we denote $\vect g(\theta;\vect X)=(g_1(\theta;\vect X),g_2(\theta;\vect X),1)$ the associated steering vector. In the following we will keep the same parameterization (\ref{fitrhophi}) that was used to fit the experimental lobe that requires only 6 coefficients. However, contrarily to this former case, there is no reference direction for the trial antenna pattern (since the antennas covariance matrix is only sensitive to the relative directions of the sources with respect to the antenna gains). A directional marker can be obtained from the coastline, which limits the admissible angular domain. We will therefore seek to optimize the antenna pattern under the form:
\be\label{parametrizationg}
\begin{split}
  g_1(\theta)&=\rho_1\cos(\theta-\theta_{1})e^{i\phi_1}H(\theta),\\
  g_2(\theta)&=\rho_2\sin(\theta-\theta_{2})e^{i\phi_2}H(\theta),
    \end{split}
\ee
where $H(\theta)$ is an angular mask delimiting the maximum opening to sea (i.e., a land mask), $H(\theta)=1$ if $\theta_{min}<\theta<\theta_{max}$ and $H(\theta)=0$ otherwise. {\color{black} Note that the method will not apply in the rare situation where there is no available land mask (that is, essentially, for HFR installed on islands with full 360 degree visibility). In that case, one needs to rely on the theoretical bearing to fix at least one angular parameter $\theta_{1}$ or $\theta_{2}$ and perform the optimization on the remaining parameters.}

\paragraph{} Record a long series of range-resolved complex Doppler spectra for each antenna (CSS files) with the standard integration time (10 min). Assuming a constant calibration over the period, this series should cover at least a few weeks and a variety of oceanic conditions (hence various current patterns). 
\paragraph{} For every nth Doppler bin above some threshold (depending on the range), calculate the antenna covariance matrix using an ensemble average over the standard duration (typically, 7 CSS files over one hour). Perform a SVD of this matrix and store the eigenvectors $\vect{V_1},\vect{V_2},\vect{V_3}$ corresponding to the 3 eigenvalues $\lambda_1,\geq \lambda_2\geq \lambda_3$. Repeat this operation for every hour of data in the database. Stack the $N$ available eigenvectors obtained with different times, range and Doppler bins in a unique large $3\times N$ matrix of $N$ vectors $(\vect{V_1^j},\vect{V_2^j},\vect{V_3^j})$. This constitutes the reference database of eigenvectors for the self-calibration process.
\paragraph{} For every set of parameters $\vect X$ characterizing the antenna pattern, evaluate a cost function ${\cali L}(\vect X)$ defined in the following way. Let us denote $\vect g(\theta;\vect X)$ the steering vector associated to this trial pattern. For every $j^{th}$ set of eigenvectors  $(\vect{V_1^j},\vect{V_2^j},\vect{V_3^j})$ and antenna parameters $\vect X$, one can evaluate a MUSIC factor:
  \be
   F(j;\vect X)=\max_\theta\left( \frac{ \norme{ \vect g(\theta;\vect X) }^2 }{\norme{ \sum_{n=3-N_s+1}^3 (\vect V_n^j\cdot \vect g(\theta;\vect X)) \vect V_n^j }^2 }\right),
   \ee
  where $N_s$ is the chosen number of sources. The main idea is that the MUSIC factors $F(j;\vect X)$ obtained by using the eigenvectors database will be amplified if the trial steering vector $g(\theta_;\vect X)$ is close to the actual steering vector at the detected DoA. When using a large database, say over one month, one can expect that the variety of encountered oceanic conditions will allow to sweep all possible bearings. To quantify the global increase of the MUSIC factors, we selected a criterion based on the median of the population.  We therefore devised the following cost function:
  \be
     {\cali L}(\vect X)=\left(\mathrm{Median}\left(F(j;\vect X)\right)\right)^{-1}
     \ee
     The Median was preferred to the Mean or the Maximum value of the population that are too sensitive to outliers and lead to a noisy and discontinuous dependence of the cost function on the trial parameters $\vect X$. The optimal set of parameters can then be estimated using a classical minimization routine with some initial guess on the parameter values. 

     {\color{black}
       Note that the method requires calculating a full population of MUSIC factors over all available triplets of eigenvectors $\vect{V_1},\vect{V_2},\vect{V_3}$ for each trial values of the 6-parameter vector $\vect X$. In our numerical trials the optimization has been obtained through a standard nonlinear least-square solver provided by the MATLAB toolbox. The computational time depends on the size of the database but remains of the order of a few minutes with a desktop computer for a monthly databasis.
     
\subsection{Comparison with  Friedlander and Weiss method}
The present self-calibration method shares common ideas with a family of techniques that were proposed three decades ago in the signal processing literature to infer the complex antenna gains of arbitrary arrays from the eigen-structure analysis of the received signal (see \cite{friedlander1988eigenstructure,weiss1990eigenstructure} for the original method and \cite{tuncer2009classical} for a modern review). In this approach, henceforth referred to as Friedlander and Weiss (FW) method, the individual phase and amplitude corrections on each antenna are sought by minimizing a cost function related to a MUSIC factor. However, there are some major differences both in the antenna configuration and in the algorithm that prevents it to being adapted to the present situation. First, the method is designed for basic phased arrays for which the antenna calibration is reduced to a constant gain correction on each antenna and is not adapted to compact arrays such as Seasonde that rely on the identification of a specific antenna pattern. Second, the algorithm assumes a fixed and small number $N$ of stationary sources, which must be smaller than the number $M$ of antennas in the array. Hence the optimization of the antenna parameters is based on only one estimation of the signal covariance matrix (that is, a unique set of eigenvectors) whereas our method uses an arbitrary large number of sources corresponding to the many occurrences of DoA that are found at different ranges and times due to the diversity of radial current spatial patterns. In the FW method, the fixed number and position of sources is compensated for by using the $N$ highest peaks of the MUSIC function instead of a single MUSIC factor. The resulting constraint on the cost function is sufficient to invert the $M$ antenna parameters. This is not applicable to SeaSonde that is limited to a maximum of 2 sources. Note that this would also be problematic with HFR phased arrays as this would require a good evaluation of the signal eigenvalues and an unambiguous separation with the noise eigenvalues, a condition which is rarely granted due to the actual noise level. Yet, the FW method was applied recently to phased-array oceanographic radars (\cite{Zhao_JAOT15}) and shown to improve the estimation of radial surface current. However, we could not find in this paper the technical details indicating how the number of sources was selected, an important issue for the performance of the algorithm.}

\subsection{Application to the CODAR station}
The aforementioned self-calibration procedure was applied to the SeaSonde data using one full month of record (May, 2019). The six parameters of the cost function were optimized using the minimization toolbox of Matlab and the ideal pattern parameter as an initial guess. The obtained antenna patterns are summarized in Table \ref{tableparam}, together with a recap of the parameter values for the ideal and experimental lobes. Note that the estimated as well as measured antenna patterns are far from satisfying the expected quadrature relation between loop 1 and loop 2, judging from the significantly different values obtained for the bearings $\theta_1$ and $\theta_2$ with a discrepancy of about 15 degrees.

Figure \ref{cartes3methodes} shows one example of radial map obtained with the self-calibrated antenna patterns and its comparison with those obtained with the ideal and experimental antenna patterns. The small differences in the amplitude and bearing antenna parameters result in an overall rotation and dilation of the map. It also illustrates the advantage of using temporal stacking. As seen, this technique significantly improves the coverage of the map of radial current and prevents from using interpolation techniques that could introduce artifacts in the distribution. Note, however, that the azimuthal resolution has not been increased and is still limited by the geometry of the antenna receive array.
  
 \subsection{ {\color{black} Is the antenna pattern stable in time ?}}
        
        {\color{black} In the previous comparison we have shown non contemporary estimates of the experimental (March 2020) and self-calibrated (May 2019) patterns, the choice of the latter period being driven
          by the date of drifter launch. Now it is possible that the actual antenna lobes have evolved over time (here, 10 months), which would explain part of the observed difference. To check this hypothesis we also performed the self-calibration using one month of data in March 2020. The obtained parameters are shown in the last line of Table \ref{tableperf}. The loop amplitude ratios are found closer to the experimentally measured values with a marked unbalance between antenna 1 and 2. The differences with the experimental bearings and phases could be due to the inaccuracy in measuring the latter (see Figure \ref{figloopamplitude}) and its dependence to the degree of angular smoothing. Nevertheless, judging from the clear trend of the amplitude ratio, it is likely that the actual antenna pattern has been subject to some change over time, which further justifies the use of updated self-calibration when extracting the surface current maps. }
        
\subsection{One or two sources ?}
The number of sources that can be used in the MUSIC DF algorithm is, at most, one unity less than the number of receive antennas. Prescribing the optimal number of sources has a significant impact on the accuracy of the estimation (e.g. \cite{kirincich_JAOT19}]. In the case of the CODAR SeaSonde that has only 3 antennas, the possible number of sources is limited to 1 or 2. In principle, using 2 sources instead of 1 provides a better estimation of the spatial structure of the radial current, in particular in non-trivial situations when the latter has meandering or vortexes leading to multiple DoA for a given radial current value. However, using a two-source DF is very demanding in terms of accurate knowledge of the antenna pattern and level of SNR for the Doppler bins as it requires an unambiguous identification of the second largest signal eigenvalue. As a secondary MUSIC peak is most often observed in practice, prescribing 2 sources can lead to a spurious peak detection and a wrong allocation of the DoA. This was clearly the case with the SeaSonde data of Saint-Jean Cap Ferrat, where the utilization of a two-source MUSIC research resulted in many errors in the estimation of the DoA and produced many outliers in the surface current map. We therefore decided to perform the whole azimuthal processing with one single source, which is less accurate but more robust than two sources in this case. Similarly, the self-calibration of the antenna pattern was realized using a single source in a region of strong SNR (range cells 7 to 17). {\color{black}
The lesser performance of the 2-source self-calibration was confirmed by the drifter comparisons (Table \ref{tableperf} in Section \ref{sec:drifter}).}
    
\begin{figure*}[htbp]
\begin{minipage}[c]{0.5\textwidth}
a) \includegraphics[scale=0.425]{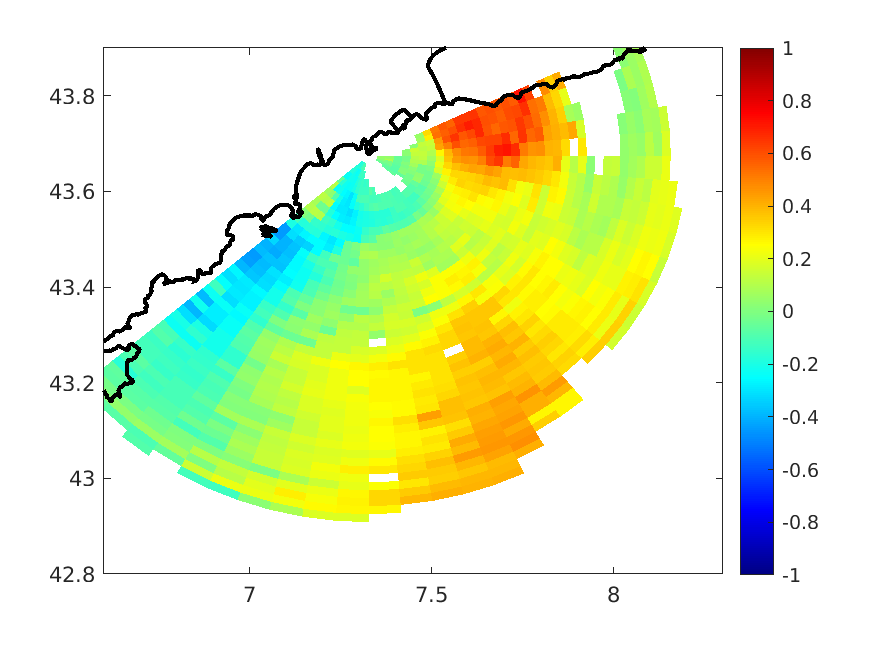}
\end{minipage} \begin{minipage}[c]{0.5\textwidth}
b) \includegraphics[scale=0.425]{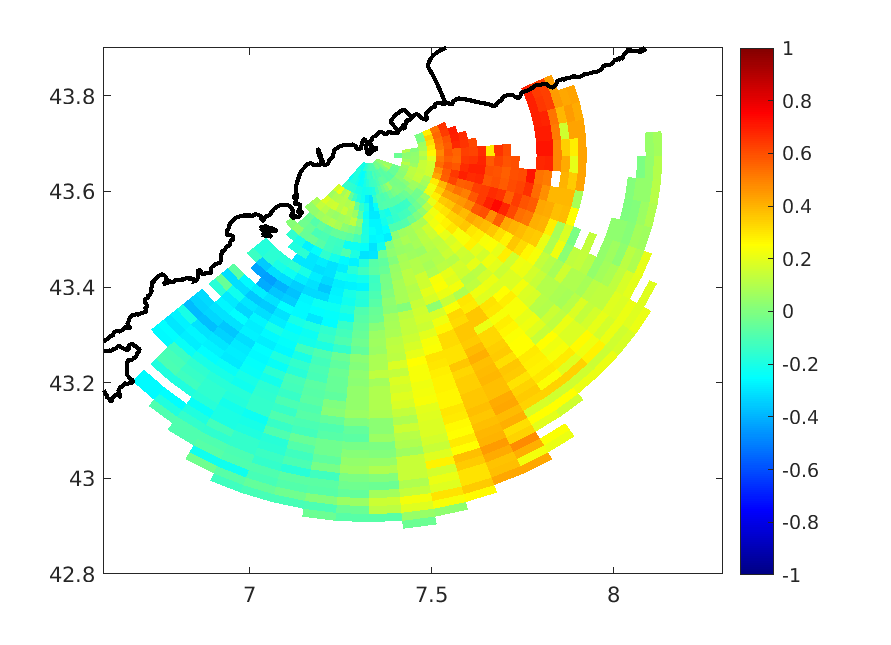}  \end{minipage}
\begin{minipage}[c]{0.5\textwidth}
c) \includegraphics[scale=0.425]{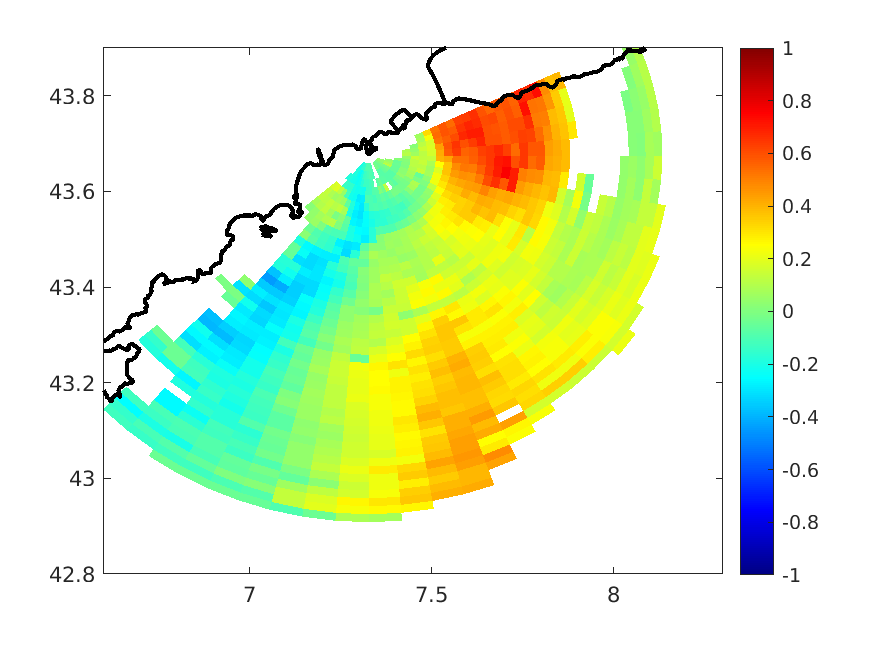}
\end{minipage}\begin{minipage}[c]{0.5\textwidth}
    d) \includegraphics[scale=0.425]{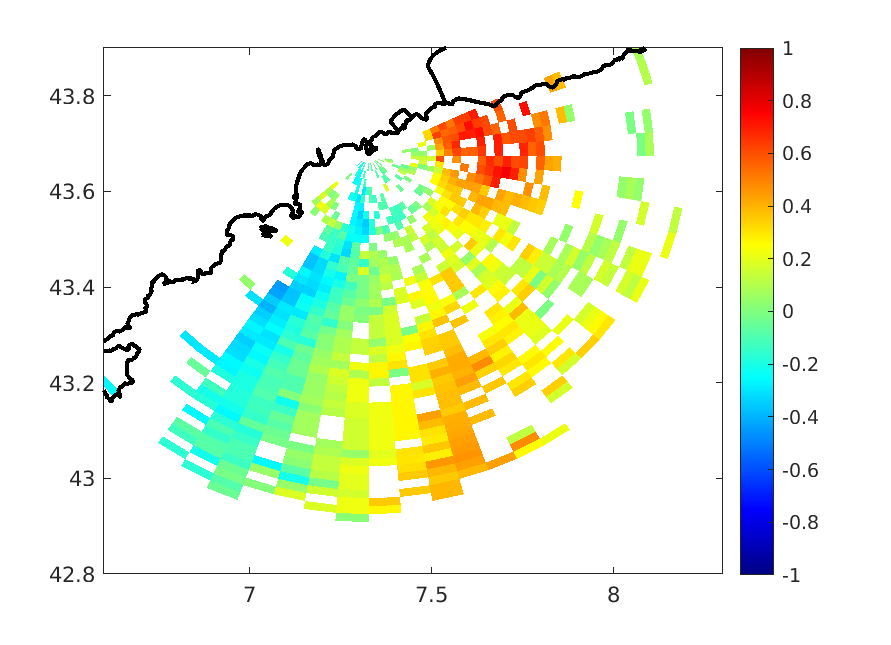}
\end{minipage}
 \caption{Examples of radial current maps obtained at the Ferrat station using 1 hour of record  (May, 13, 2019, 16:00 UTC) and the (a) ideal, (b) experimental and self-calibrated (c) antenna pattern. Temporal stacking and one single source have been used in the DoA algorithm. The effect of stacking is illustrated in (d), where the radial map obtained with the self-calibrated antenna pattern (c) is recalculated without temporal stacking. \label{cartes3methodes}}
\end{figure*}
     
\begin{table*}[htbp]\centering
\center
    \begin{tabular}{|c|c|c|c|c|c|c|}
  \hline
  & $\rho_1$ & $\rho_2$ & $\theta_1$  & $\theta_2$ &  $\phi_1$ & $\phi_2$\\
  \hline
Ideal & 1 & 1 & 212 & 212  & 0 & 0 \\
  \hline
Experimental  (March 2020) & 2.9 & 1.9 & 225 & 210  &-28 &-41 \\
  \hline
  Self-Calibrated (May 2019) & 1.43 & 1.43 & 225.6 & 205.1 & -9.9 & -20.9\\
  \hline
\color{black}  Self-Calibrated (March 2020) & 2.2 & 1.3 & 210.5 & 201.7 & 23.7 & 32.6\\
\hline
\end{tabular}
    \color{black}\caption{Estimation of the antenna pattern parameters for the ideal, experimental and self-calibrated lobes at the CODAR station Saint-Jean Cap Ferrat. The angles are given in degrees NCW.\label{tableparam}}
\end{table*}
   
\section{Comparison with drifters\label{sec:drifter}}

\begin{figure}
  \centering
   \hspace{-0.2cm}  \includegraphics[scale=0.6]{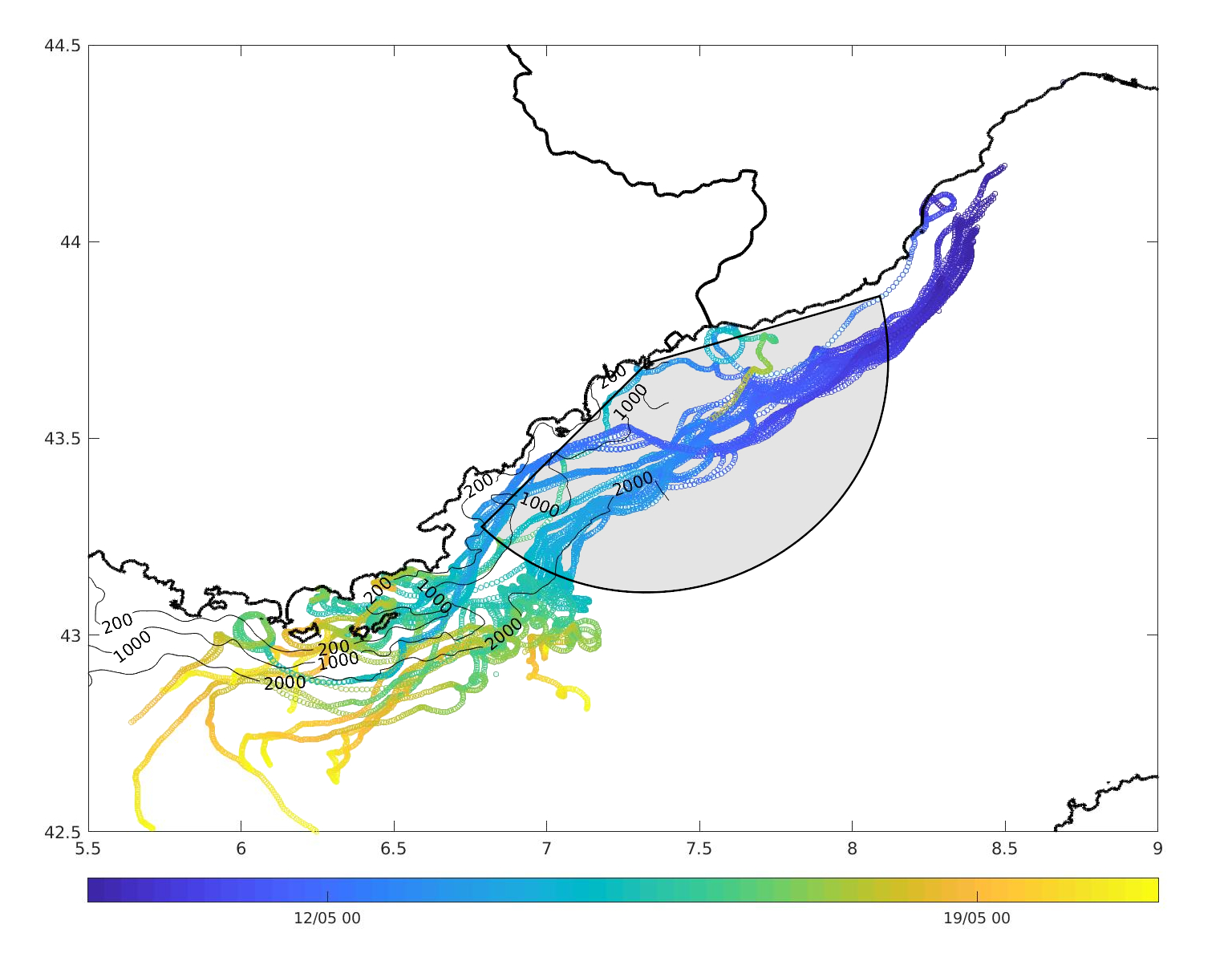}
\caption{Drifters trajectories from May, 10 to May, 20, 2019. The color scale indicates the running date. The typical radar footprint of the Ferrat station is shown with the black arc of circle.\label{driftermap}}
\end{figure}

{\color{black}To assess the quality of an antenna calibration, it is customary to compare the HFR products with {\it in situ} instruments such as drifters or ADCPs (e.g. \cite{paduan_JOE06,Liu_JAOT10,kirincich_JAOT12,Zhao_JAOT15}).} To evaluate the respective merits of the different calibration methods, we here compared the HFR radial surface currents with drifter measurements obtained during an experiment conducted in May, 2019. To this aim, we re-calculated the self-calibrated antenna pattern using the corresponding month of data. An example of a typical radial current map seen from this station is shown in Figure \ref{cartes3methodes}b. The drifter set, used as ground truth, is part of the dataset collected under the IMPACT EU project (\cite{impact_impact_2016}) aiming at providing tools and guidelines to combine the conservation of the marine protected areas with the development of port activities in the transboundary areas in between France and Italy. The deployment includes 40 CARTHE-type drifters, biodegradable up to 85\% (\cite{novelli_biodegradable_2017}), released in the open sea in front of the Gulf of La Spezia (Italy) on May 2, 2019 (\cite{berta_m_drifter_2020}). The deployment strategy consisted of releasing 35 drifters on a regular grid (about 6 km side and 1 km step, with some 500 m nesting) in a time window of about 2-3 hours. 5 more drifters were released in line after passing by the Portovenere Channel, while exiting the Gulf of La Spezia. The nominal drifter transmission rate was 5 min, GPS transmitters nominal accuracy was within 10 m, and drifter positions were pre-processed to remove outliers.
Most of the drifters followed the prevailing current pattern in the area, represented by the NC, and moved along the continental margin from the North East to the South West in agreement with the general cyclonic circulation of the NW Mediterranean basin. Most of the drifters reached the area of interest, around May 10, and 35 trajectories were exploitable for the HFR currents assessment in between May 10 and May 20, wherever a non-void intersection was obtained with the HFR surface current estimates. In order to compare with HFR data, drifter positions were interpolated on the HFR time vector and drifter velocities were projected onto the radial direction of the antenna pattern. The surface layer sampled by CARTHE drifters covers the first 60 cm of the water column (\cite{novelli_biodegradable_2017}). This is consistent with the water layer measured by HFR since velocities are vertically averaged down to a depth estimated as $\lambda/8\pi$, where $\lambda$ is the transmitted radar wavelength (\cite{stewart_hf_1974}), which corresponds approximately to a layer of about 88 cm for the considered HFR operating frequency 13.5 MHz. CARTHE drifters have low windage, thanks to the flat toroidal floater. In addition, their motion is not affected by rectification caused by wind waves, thanks to the flexible tether connecting the floater and the drogue; at last, their slip velocity is less than 0.5\% for winds up to 10 m/s (\cite{novelli_biodegradable_2017}). Some of these CARTHE drifters reached the area of Toulon in the stream of the NC around May, 19, 2019 (see Figure \ref{driftermap}) and their data were already compared with the radar estimates from the local HFR stations (a WERA bistatic network at 16.15 operating frequency (see \cite{guerin_multistatic_2019})). An excellent agreement was found for the corresponding radial currents (\cite{dumas_multistatic_2020}), thereby confirming that the CARTHE are very relevant instruments to calibrate HFR surface current measurements.

\begin{figure*}
  \centering
\includegraphics[scale=0.32]{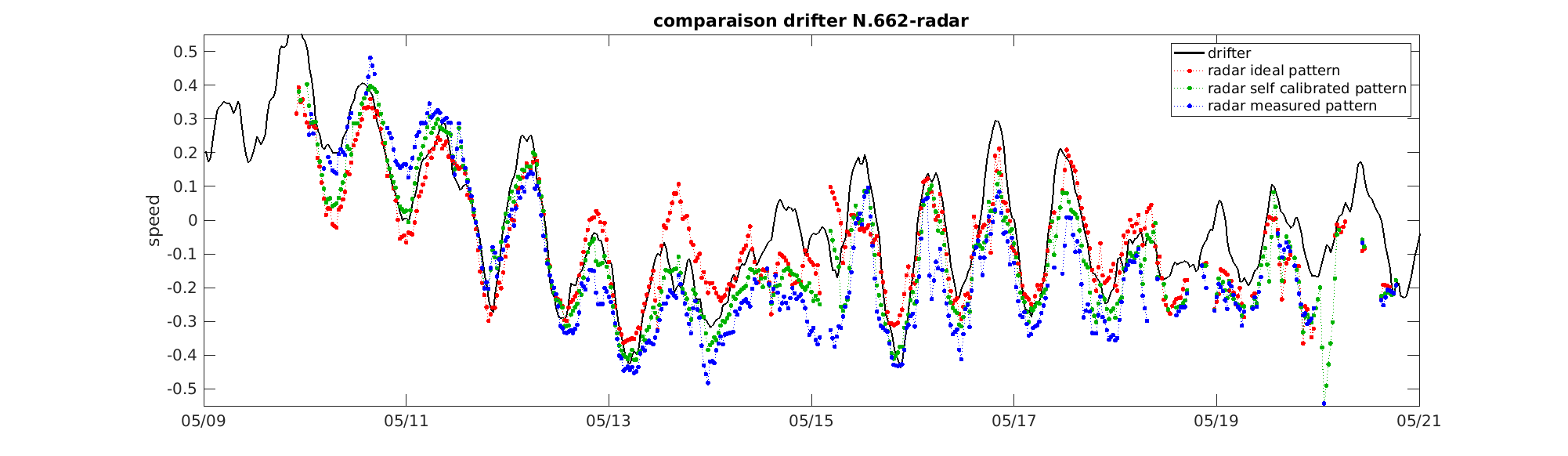}
  \caption{Comparisons of HFR derived radial surface current (dots) with drifter measurements (black solid line) for one complete trajectory. The different colors (red, green and blue) refer to the different antenna patterns that have been used (ideal, experimental and self-calibrated, respectively).\label{goodtrajectories}}
\end{figure*}

For the purpose of comparison, the drifter velocity were projected onto the local radial seen from the HFR site and the hourly HFR measurements were interpolated both in space and time to match the drifter locations. We inspected the HFR velocities obtained with each of the 3 calibration methods along the individual drifter trajectories. {\color{black} The DoA method was run as presented in Section \ref{sec:DoA} using temporal stacking and assuming one single source.} Figure \ref{goodtrajectories} shows an example of such comparison for one trajectory that remained entirely within the radar coverage over the duration of the experiment. As seen, the dominant inertial oscillations are well captured by the HFR estimates for all types of antenna calibration with the self-calibration method being at first sight best performing. However, some non-negligible errors are visible in the magnitude of the peaks around May, 10 and May, 15, 2019.
To quantify the performances of the different calibration techniques we calculated the main statistical parameters of the overall dataset. By concatenating the time series from the 40 trajectories, we obtained an ensemble of 8090 simultaneous values of HFR and drifter radial currents over the entire experiment ($U_{radar}$ and $U_{drifter}$, respectively). From this ensemble of values we calculated the bias ($\langle U_{radar}-U_{drifter}\rangle$), the Root Mean Square Error (RMSE=$(\langle [U_{radar}-U_{drifter}]^2\rangle)^{1/2}$) and the correlation coefficient of the linear regression for each of the 3 calibration methods. The results are summarized in Table \ref{tableperf}. As seen, using the experimental antenna patterns only marginally improves the accuracy for the estimated radial currents when compared to the ideal pattern. However, the use of the self-calibrated antenna pattern allows for a noticeable decrease of RMSE by ~2.3 cm/s with respect to the ideal pattern.

{\color{black} Another useful classical test (e.g. \cite{Liu_JAOT10}) to evaluate the correctness of the antenna bearing is to compare the drifter measurements with the HFR currents after performing a small azimuthal rotation of the latter. As the HFR current map was realized using a 5 degree angular bin (which was found to be a good trade-off between the azimuthal accuracy of the current map and the required size to avoid lacunary patterns in the DoA algorithm) we could only evaluate the effect of shifting the HFR cells by multiples of $\pm 5$ degrees. Table \ref{tablestacking} shows the impact of the bearing offset on the comparison performance using the 3 types of antenna patterns. Judging from the optimal performance (that is, minimizing the RMSE and maximizing the CC), it appears that the self-calibrated pattern has a small angle offset (no more than 5 degrees) while the experimental pattern is found with a slightly larger offset (about 10 degrees).
  }

\begin{figure}
\hspace{-0.25cm} \includegraphics[scale=0.45]{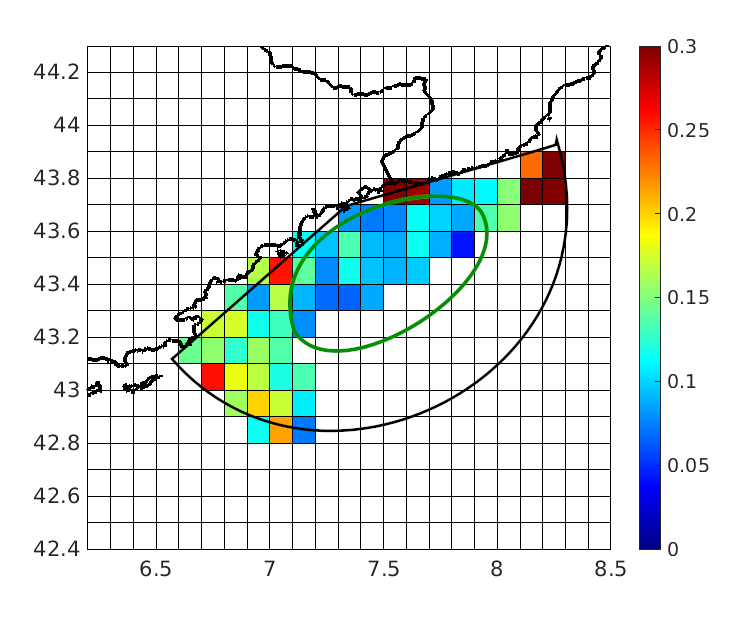}
  
\caption{RMSE (in m/s) between self-calibrated HFR currents and drifter measurements calculated with all available sampling points along the 40 trajectories and binned by 0.1 deg bins in latitude and longitude. The black arc of circle shows the typical radar footprint while the green ellipse eliminates the central region where the error is smallest.\label{errormap}}
\end{figure}

To better identify the area where the HFR estimates are of lesser quality, we calculated the RMSE in restricted areas corresponding to 0.1 degree bins of longitude and latitude and retained only the grid cells where at least 10 sample points were available (out of 8090). Figure \ref{errormap} shows the obtained RMSE in colorscale, superimposed to the map of radar coverage. From this refined analysis it turns out that the accuracy of the HFR estimation is greatly improved in a central zone that can be coarsely delimited by a square area comprised between 7.3-7.9 longitude and 43.3-43.7 latitude. The RMSE obtained by considering all data in this central zone is found to about 10 cm/s when using the self-calibrated antenna pattern, which is about 3 cm/s lower than the corresponding RMSE calculated with the ideal or measured antenna pattern. Given the magnitude of the radial surface current (up to 1 m/s) this is a good performance.

\begin{table}[htbp]\centering
  \center
\begin{tabular}{|c|c|c|c|c|}
  \hline
  & bias & RMSE & CC\\
  \hline
  ideal & 4.8 (8.6)& 15.8 (13.1) &0.85 (0.91)\\
   \hline
   measured & 7.9 (4.2) & 15.3 (13.4) & 0.88 (0.97)  \\
    \hline
   self-calibrated& 7.1 (3.6) & 13.5 (9.7) &0.90 (0.92)\\
   \hline
\end{tabular}
\caption{Overall statistical performances of the HFR radial current measurements when compared to the 40 drifters. The bias and RMSE are given in cm/s and CC is the correlation coefficient. In parenthesis, the same statistical performances on the reduced area comprised between 7.3-7.9 longitude and 43.3-43.7 latitude. \label{tableperf}}
\end{table}

\begin{table*}[htbp]\centering
  \center
   \small
\begin{tabular}{|l|l|c|c|c|c|c|c|c|}
   \hline
 Pattern & offset & -15 & -10 & -5 & 0 & 5 & 1 0 & 15 \\
  \hline
  Ideal with &  RMSE& 21.6 (21.5)& 19.5 (18.8)& 17.2 (16.7)& 15.8 ({\bf 14.9})& {\bf 15.3} (15.2)& 15.5 (15.8)& 16.3 (16.6)\\
    (w/o) stacking & CC & 0.79 (0.82)& 0.82 ({\bf 0.86})& 0.84 (0.85)& {\bf 0.85} (0.85)& 0.84 (0.85)& 0.83 (0.82)& 0.81 (0.81)\\
   \hline
Experimental (3/2020)& RMSE & 18.7 (19.4)& 17.4 (18.2)& 16.0 (17.7)& 15.3 (15.5)& 14.4 (15.0)& {\bf 14.1} ({\bf 14.6})& 14.3 (15.5) \\
with (w/o) stacking & CC &{\bf 0.90} ({\bf 0.90})&{\bf  0.90} ({\bf 0.90})& {\bf 0.90} (0.88)& 0.88 (0.87)& 0.84 (0.85)& 0.85 (0.83)& 0.81 (0.79)\\
    \hline
Self-calibrated (5/2019) & RMSE & 20.6 (21.0)& 18.0 (17.7)& 15.6 (15.6)& 13.5 (13.0)& {\bf 12.5} ({\bf 12.4})& 13.6 (13.7)& 15.0 (16.1)\\
 with (w/o) stacking)& CC & 0.86 (0.87)& 0.89 (0.90)& {\bf 0.90} ({\bf 0.91})& {\bf 0.90} ({\bf 0.91})& {\bf 0.90} (0.88)& 0.87 (0.85)& 0.83 (0.81)\\
    \hline
Same for 2 sources & RMSE & 22.2 (23.4)& 20.1 (21.0)& 18.2 (18.5)& 15.9 (16.5)& 13.8 (13.9)& {\bf 12.8} ({\bf 12.3})& 13.1 (13.2)\\
with (w/o) stacking & CC & 0.83 (0.84)& 0.88 (0.88)& 0.89 (0.90)& 0.90 (0.91)& {\bf 0.91} ({\bf 0.91})& 0.90 (0.90)& 0.88 (0.86)\\
\hline

\end{tabular}

\normalsize

\caption{{\color{black}Influence of the chosen pattern, the bearing offset and the stacking on the overall performances of the HFR currents when compared to the drifter measurements. The RMSE is given in m/s. The optimal bearing for the RMSE or the CC are highlighted in bold.\label{tablestacking}}}
\end{table*} 

\color{black}\normalsize

\section{Conclusion}

We have presented in detail the first results of one SeaSonde CODAR station in the region of Nice together with the specific processing techniques that have been developed in the laboratory. An original self-calibration technique of the antenna patterns has been applied and allows to improve the quality of the HFR radial currents. The accuracy of the latter has been assessed with the help of a set of drifters measurements and the RMSE has been shown to be diminished by 2-3 cm/s with respect to the HFR estimations based on the ideal or experimental antenna patterns. The strong point of this self-calibration procedure is that it does not require any extra ship campaign nor external sources and is purely based on the analysis of the recorded time series that can be updated at will. The results pertaining to the companion HFR station recently installed in Menton will be presented in a subsequent work, together with the vector reconstruction of the radial currents.

\acknowledgments \small{The CODAR SeaSonde sensors of the MIO were jointly purchased by several French Research institutions (Universit\'e de Toulon, CNRS, IFREMER). Their maintenance has been supported by the national observation service ``MOOSE'' and the EU Interreg Marittimo projects SICOMAR-PLUS and IMPACT. Special thanks go to the office of ``Phares et Balises DIRM M{\'e}diterran{\'e}e'' for hosting our installation in Saint-Jean Cap Ferrat; the Institut de la Mer de Villefranche for lending the Pelagia boat; our colleagues D. Mallarino, J-L. Fuda, T. Missamou and R. Chemin for assisting with installation; our colleague Marcello Magaldi for the IMPACT drifters campaign; Marcel Losekoot (UC Davis) and Brian Emery (UC Santa Barbara) for providing software to read the SeaSonde time series. {\color{black}We are also grateful to one anonymous reviewer for indicating the Friedlander and Weiss method for self-calibration of arrays.}}
\datastatement
Data available on demand.

\bibliographystyle{ametsoc2014}

\begin{thebibliography}{40}
\providecommand{\natexlab}[1]{#1}
\providecommand{\url}[1]{\texttt{#1}}
\renewcommand{\UrlFont}{\rmfamily}
\providecommand{\urlprefix}{URL }
\expandafter\ifx\csname urlstyle\endcsname\relax
  \providecommand{\doi}[1]{doi:\discretionary{}{}{}#1}\else
  \providecommand{\doi}{doi:\discretionary{}{}{}\begingroup
  \urlstyle{rm}\Url}\fi
\providecommand{\eprint}[2][]{\url{#2}}

\bibitem[{Bellomo et~al.(2015)}]{bellomo_toward_2015}
Bellomo, L., and Coauthors, 2015: Toward an integrated {HF} radar network in
  the {Mediterranean} {Sea} to improve search and rescue and oil spill
  response: the {TOSCA} project experience. \textit{Journal of Operational
  Oceanography}, \textbf{8~(2)}, 95--107, \doi{10.1080/1755876X.2015.1087184},
  number: 2.

\bibitem[{Berta et~al.(2020)Berta, Sciascia, Magaldi,, and
  Griffa}]{berta_m_drifter_2020}
Berta, M., R.~Sciascia, M.~Magaldi, and A.~Griffa, 2020: Drifter deployment in
  the framework of the {IMPACT} ({Port} impacts on marine protected areas:
  transnational cooperative actions) project. {SEANOE}, CNR ISMAR.
  \urlprefix\url{https://doi.org/10.17882/72369}.

\bibitem[{Berta et~al.(2018)}]{berta_OS18}
Berta, M., and Coauthors, 2018: Wind-induced variability in the northern
  current (northwestern mediterranean sea) as depicted by a multi-platform
  observing system. \textit{Ocean Science}, \textbf{14~(4)}, 689--710,
  \urlprefix\url{https://doi.org/10.5194/os-14-689-2018}.

\bibitem[{Coppola et~al.(2018)Coppola, Legendre, Lefevre, Prieur, Taillandier,,
  and Riquier}]{coppola_PO18}
Coppola, L., L.~Legendre, D.~Lefevre, L.~Prieur, V.~Taillandier, and E.~D.
  Riquier, 2018: Seasonal and inter-annual variations of dissolved oxygen in
  the northwestern mediterranean sea (dyfamed site). \textit{Progress in
  Oceanography}, \textbf{162}, 187--201.

\bibitem[{Coppola et~al.(2019)Coppola, Raimbault, Mortier,, and
  Testor}]{coppola_EOS19}
Coppola, L., P.~Raimbault, L.~Mortier, and P.~Testor, 2019: Monitoring the
  environment in the northwestern mediterranean sea. \textit{Eos, Transactions
  American Geophysical Union}, \textbf{100}.

\bibitem[{Dumas et~al.(2020)Dumas, Gramoull{\'e}, Gu{\'e}rin, Molcard,
  Ourmieres,, and Zakardjian}]{dumas_multistatic_2020}
Dumas, D., A.~Gramoull{\'e}, C.-A. Gu{\'e}rin, A.~Molcard, Y.~Ourmieres, and
  B.~Zakardjian, 2020: Multistatic estimation of high-frequency radar surface
  currents in the region of {Toulon}. \textit{Ocean Dynamics},
  \textbf{70~(12)}, 1485--1503, number: 12 Publisher: Springer.

\bibitem[{Dumas and Gu{\'e}rin(2020)Dumas, and
  Gu{\'e}rin}]{dumas_self-calibration_2020}
Dumas, D., and C.-A. Gu{\'e}rin, 2020: Self-calibration and antenna grouping
  for bistatic oceanographic {High}-{Frequency} {Radars}. \textit{arXiv
  preprint arXiv:2005.10528}.

\bibitem[{Emery and Washburn(2019)Emery, and Washburn}]{emery_JAOT19}
Emery, B., and L.~Washburn, 2019: Uncertainty estimates for seasonde hf radar
  ocean current observations. \textit{Journal of Atmospheric and Oceanic
  Technology}, \textbf{36~(2)}, 231 -- 247, \doi{10.1175/JTECH-D-18-0104.1},
  \urlprefix\url{https://journals.ametsoc.org/view/journals/atot/36/2/jtech-d-18-0104.1.xml}.

\bibitem[{Emery et~al.(2014)Emery, Washburn, Whelan, Barrick,, and
  Harlan}]{emery_JAOT14}
Emery, B.~M., L.~Washburn, C.~Whelan, D.~Barrick, and J.~Harlan, 2014:
  Measuring antenna patterns for ocean surface current hf radars with ships of
  opportunity. \textit{Journal of Atmospheric and Oceanic Technology},
  \textbf{31~(7)}, 1564--1582.

\bibitem[{Fernandez et~al.(2003)Fernandez, Vesecky,, and
  Teague}]{fernandez_IGARSS03}
Fernandez, D.~M., J.~Vesecky, and C.~Teague, 2003: Calibration of hf radar
  systems with ships of opportunity. \textit{IGARSS 2003. 2003 IEEE
  International Geoscience and Remote Sensing Symposium. Proceedings (IEEE Cat.
  No. 03CH37477)}, IEEE, Vol.~7, 4271--4273.

\bibitem[{Fernandez et~al.(2006)Fernandez, Vesecky,, and
  Teague}]{fernandez_JOE06}
Fernandez, D.~M., J.~F. Vesecky, and C.~Teague, 2006: Phase corrections of
  small-loop hf radar system receive arrays with ships of opportunity.
  \textit{IEEE Journal of Oceanic Engineering}, \textbf{31~(4)}, 919--921.

\bibitem[{Flores-Vidal et~al.(2013)Flores-Vidal, Flament, Durazo, Chavanne,,
  and Gurgel}]{flores_JAOT13}
Flores-Vidal, X., P.~Flament, R.~Durazo, C.~Chavanne, and K.-W. Gurgel, 2013:
  High-frequency radars: Beamforming calibrations using ships as reflectors.
  \textit{Journal of Atmospheric and Oceanic Technology}, \textbf{30~(3)},
  638--648.

\bibitem[{Friedlander and Weiss(1988)Friedlander, and
  Weiss}]{friedlander1988eigenstructure}
Friedlander, B., and A.~J. Weiss, 1988: Eigenstructure methods for direction
  finding with sensor gain and phase uncertainties. \textit{ICASSP-88.,
  International Conference on Acoustics, Speech, and Signal Processing}, IEEE,
  2681--2684.

\bibitem[{Gu{\'e}rin et~al.(2019)Gu{\'e}rin, Dumas, Gramoull{\'e}, Quentin,
  Saillard,, and Molcard}]{guerin_multistatic_2019}
Gu{\'e}rin, C.-A., D.~Dumas, A.~Gramoull{\'e}, C.~Quentin, M.~Saillard, and
  A.~Molcard, 2019: The multistatic {HF} radar network in {Toulon}.
  \textit{{IEEE} {Radar} 2019 {Conference}}, IEEE.

\bibitem[{IMPACT(2016)}]{impact_impact_2016}
IMPACT, 2016: \textit{{IMpact} of {Ports} on marine protected areas:
  {Cooperative} {Cross}-{Border} {Actions}}. www.impact-maritime.eu.

\bibitem[{Kirincich(2017)}]{kirincich_JAOT17}
Kirincich, A., 2017: Improved detection of the first-order region for
  direction-finding hf radars using image processing techniques.
  \textit{Journal of Atmospheric and Oceanic Technology}, \textbf{34~(8)}, 1679
  -- 1691, \doi{10.1175/JTECH-D-16-0162.1},
  \urlprefix\url{https://journals.ametsoc.org/view/journals/atot/34/8/jtech-d-16-0162.1.xml}.

\bibitem[{Kirincich et~al.(2019)Kirincich, Emery, Washburn,, and
  Flament}]{kirincich_JAOT19}
Kirincich, A., B.~Emery, L.~Washburn, and P.~Flament, 2019: Improving surface
  current resolution using direction finding algorithms for multiantenna
  high-frequency radars. \textit{Journal of Atmospheric and Oceanic
  Technology}, \textbf{36~(10)}, 1997--2014.

\bibitem[{Kirincich et~al.(2012)Kirincich, De~Paolo,, and
  Terrill}]{kirincich_JAOT12}
Kirincich, A.~R., T.~De~Paolo, and E.~Terrill, 2012: Improving hf radar
  estimates of surface currents using signal quality metrics, with application
  to the mvco high-resolution radar system. \textit{Journal of Atmospheric and
  Oceanic Technology}, \textbf{29~(9)}, 1377--1390.

\bibitem[{Laws et~al.(2010)Laws, Paduan,, and Vesecky}]{Laws_JAOT10}
Laws, K., J.~D. Paduan, and J.~Vesecky, 2010: Estimation and assessment of
  errors related to antenna pattern distortion in codar seasonde high-frequency
  radar ocean current measurements. \textit{Journal of Atmospheric and Oceanic
  Technology}, \textbf{27~(6)}, 1029--1043.

\bibitem[{Lipa et~al.(2006)Lipa, Nyden, Ullman,, and Terrill}]{lipa_JOE06}
Lipa, B., B.~Nyden, D.~S. Ullman, and E.~Terrill, 2006: Seasonde radial
  velocities: Derivation and internal consistency. \textit{IEEE Journal of
  Oceanic Engineering}, \textbf{31~(4)}, 850--861,
  \doi{10.1109/JOE.2006.886104}.

\bibitem[{Liu et~al.(2014)Liu, Weisberg,, and Merz}]{Liu_JAOT14}
Liu, Y., R.~H. Weisberg, and C.~R. Merz, 2014: Assessment of codar seasonde and
  wera hf radars in mapping surface currents on the west florida shelf.
  \textit{Journal of Atmospheric and Oceanic Technology}, \textbf{31~(6)},
  1363--1382.

\bibitem[{Liu et~al.(2010)Liu, Weisberg, Merz, Lichtenwalner,, and
  Kirkpatrick}]{Liu_JAOT10}
Liu, Y., R.~H. Weisberg, C.~R. Merz, S.~Lichtenwalner, and G.~J. Kirkpatrick,
  2010: Hf radar performance in a low-energy environment: Codar seasonde
  experience on the west florida shelf. \textit{Journal of Atmospheric and
  Oceanic Technology}, \textbf{27~(10)}, 1689 -- 1710,
  \doi{10.1175/2010JTECHO720.1},
  \urlprefix\url{https://journals.ametsoc.org/view/journals/atot/27/10/2010jtecho720_1.xml}.

\bibitem[{Manso-Narvarte et~al.(2020)Manso-Narvarte, Fredj, Jord{\`a}, Berta,
  Griffa, Caballero,, and Rubio}]{manso_OS20}
Manso-Narvarte, I., E.~Fredj, G.~Jord{\`a}, M.~Berta, A.~Griffa, A.~Caballero,
  and A.~Rubio, 2020: 3d reconstruction of ocean velocity from high-frequency
  radar and acoustic doppler current profiler: a model-based assessment study.
  \textit{Ocean Science}, \textbf{16~(3)}, 575--591.

\bibitem[{Mantovani et~al.(2020)}]{mantovani2020best}
Mantovani, C., and Coauthors, 2020: Best practices on high frequency radar
  deployment and operation for ocean current measurement. \textit{Frontiers in
  Marine Science}, \textbf{7}, 210.

\bibitem[{MOOSE(2008-)}]{MOOSE}
MOOSE, 2008-: \textit{Mediterranean Ocean Observing System for the
  Environment}. https://www.moose-network.fr/fr.

\bibitem[{Novelli et~al.(2017)Novelli, Guigand, Cousin, Ryan, Laxage, Dai,
  Haus,, and {\"O}zg{\"o}kmen}]{novelli_biodegradable_2017}
Novelli, G., C.~Guigand, C.~Cousin, E.~Ryan, J.~Laxage, H.~Dai, K.~Haus, and
  T.~{\"O}zg{\"o}kmen, 2017: A {Biodegradable} {Surface} {Drifter} for {Ocean}
  {Sampling} on a {Massive} {Scale}. \textit{J. Atmos. Oceanic Technol.},
  \textbf{34}, 2509--2532, \doi{10.1175/JTECH-D-17-0055.1}.

\bibitem[{Paduan et~al.(2006)Paduan, Kim, Cook,, and Chavez}]{paduan_JOE06}
Paduan, J.~D., K.~C. Kim, M.~S. Cook, and F.~P. Chavez, 2006: Calibration and
  validation of direction-finding high-frequency radar ocean surface current
  observations. \textit{IEEE Journal of Oceanic Engineering}, \textbf{31~(4)},
  862--875, \doi{10.1109/JOE.2006.886195}.

\bibitem[{Paduan and Washburn(2013)Paduan, and Washburn}]{paduan_AnnRev13}
Paduan, J.~D., and L.~Washburn, 2013: High-frequency radar observations of
  ocean surface currents. \textit{Annual Review of Marine Science},
  \textbf{5~(1)}, 115--136, \doi{10.1146/annurev-marine-121211-172315}.

\bibitem[{Quentin et~al.(2013)}]{quentin_hf_2013}
Quentin, C., and Coauthors, 2013: {HF} radar in {French} {Mediterranean} {Sea}:
  an element of {MOOSE} {Mediterranean} {Ocean} {Observing} {System} on
  {Environment}. \textit{Ocean \& {Coastal} {Observation}: {Sensors} ans
  observing systems, numerical models \& information}, Nice, France, 25--30,
  \urlprefix\url{https://hal.archives-ouvertes.fr/hal-00906439}.

\bibitem[{Quentin et~al.(2014)}]{quentin_moose14}
Quentin, C.~G., and Coauthors, 2014: High frequency surface wave radar in the
  french mediterranean sea: an element of the mediterranean ocean observing
  system for the environment. \textit{7th EuroGOOS Conference, Oct 2014,
  Lisboa, Portugal, ISBN 978-2-9601883-1-8}, 111--118,
  \urlprefix\url{https://hal. archives-ouvertes.fr/hal-01131489}.

\bibitem[{Schmidt(1986)}]{schmidtAP86}
Schmidt, R., 1986: Multiple emitter location and signal parameter estimation.
  \textit{IEEE transactions on antennas and propagation}, \textbf{34~(3)},
  276--280.

\bibitem[{Schr{\"o}der and L{\"u}then(2009)Schr{\"o}der, and
  L{\"u}then}]{schroder2009astrophotography}
Schr{\"o}der, K.-P., and H.~L{\"u}then, 2009: Astrophotography.
  \textit{Handbook of Practical Astronomy}, Springer, 133--173.

\bibitem[{SICOMAR-PLUS(2018)}]{sicomar_sistema_2018}
SICOMAR-PLUS, 2018: \textit{{SIstema} transfrontaliero per la sicurezza in mare
  {COntro} i rischi della navigazione e per la salvaguardia dell'ambiente
  {MARino}}. http://interreg-maritime.eu/web/sicomarplus.

\bibitem[{Stewart and Joy(1974)Stewart, and Joy}]{stewart_hf_1974}
Stewart, R.~H., and J.~W. Joy, 1974: {HF} radio measurements of surface
  currents. Elsevier, Vol.~21, 1039--1049.

\bibitem[{Tuncer and Friedlander(2009)Tuncer, and
  Friedlander}]{tuncer2009classical}
Tuncer, T.~E., and B.~Friedlander, 2009: \textit{Classical and modern
  direction-of-arrival estimation}. Academic Press.

\bibitem[{Updyke(2020)}]{APMguide}
Updyke, T., 2020: Antenna pattern measurement guide. Tech. rep., Old Dominion
  University.

\bibitem[{Washburn et~al.(2017)Washburn, Romero, Johnson, Emery,, and
  Gotschalk}]{Washburn_JAOT17}
Washburn, L., E.~Romero, C.~Johnson, B.~Emery, and C.~Gotschalk, 2017:
  Measurement of antenna patterns for oceanographic radars using aerial drones.
  \textit{Journal of Atmospheric and Oceanic Technology}, \textbf{34~(5)},
  971--981.

\bibitem[{Weiss and Friedlander(1990)Weiss, and
  Friedlander}]{weiss1990eigenstructure}
Weiss, A.~J., and B.~Friedlander, 1990: Eigenstructure methods for direction
  finding with sensor gain and phase uncertainties. \textit{Circuits, Systems
  and Signal Processing}, \textbf{9~(3)}, 271--300.

\bibitem[{Zhao et~al.(2020)Zhao, Chen, Li, Ding, Huang,, and
  Fan}]{zhao_RemSen20}
Zhao, C., Z.~Chen, J.~Li, F.~Ding, W.~Huang, and L.~Fan, 2020: Validation and
  evaluation of a ship echo-based array phase manifold calibration method for
  hf surface wave radar doa estimation and current measurement. \textit{Remote
  Sensing}, \textbf{12~(17)}, 2761.

\bibitem[{Zhao et~al.(2015)Zhao, Chen, Zeng,, and Zhang}]{Zhao_JAOT15}
Zhao, C., Z.~Chen, G.~Zeng, and L.~Zhang, 2015: Evaluating two array
  autocalibration methods with multifrequency hf radar current measurements.
  \textit{Journal of Atmospheric and Oceanic Technology}, \textbf{32~(5)},
  1088--1097.

\end{thebibliography}

\end{document}